\documentclass[aps,prd,twocolumn,nofootinbib]{revtex4-2}
\usepackage{graphicx} 
\usepackage{hyperref}
\usepackage{natbib}
\usepackage{ragged2e} 
\usepackage{amsmath}
\usepackage{color} 
\usepackage{soul} 
\begin{document}

\title{Parity solves the Strong CP problem}
\author{Ravi Kuchimanchi}
\email{raviparity@gmail.com }


\begin{abstract}
A recent paper ``What can solve the strong CP problem?" goes counter to conventional wisdom by arguing that the universe was in an initial state that combines different eigenstates of $\theta$ (of the theta vacuum of QCD), and asserts that for such an initial state, axionless solutions to the Strong CP Problem based on P and CP symmetries will not work. We discuss the nature of the theta vacuum in light of this work, and find that the conventional framing and solutions based on P and CP symmetries are on a firm footing.            

\end{abstract}
\maketitle

\section{Conventional approach}
\let\svthefootnote\thefootnote

It is well known that due to the non-trivial topology the gauge invariant vacuum of QCD is an eigenstate of the theta vacuum~\cite{BELAVIN197585, PhysRevLett.37.172, CALLAN1976334}
\begin{equation}
\left|\theta\right> = \sum_{n=-\infty}^{+\infty} e^{in\theta} \left|n\right>
\label{eq:thetavac}
\end{equation}

where $n$ is the winding number of the state $\left|n\right>$ generated by large gauge transformations of the state with $\left|n=0\right>$. 

This theta structure splits the Hilbert space into separate spaces defined by different values of $\theta$.  This is because the Hamiltonian evolution does not connect these separate spaces
\begin{equation}
    \left<\theta'| H |\theta\right> = 0,
    \label{eq:H}
\end{equation}
and because, for any observable or gauge invariant operator~\cite{Shifman_2012, doi:10.1142/S0217751X02011357}
\begin{equation}
    \left<\theta'| O |\theta\right> = 0
    \label{eq:O}
\end{equation}
for all $\theta' \neq \theta$,

Therefore we cannot transition from one theta vacuum to another by time evolution or  by making observations.  

Thus $\theta$ is a superselection parameter and states with different values of $\theta$ belong to different Hilbert spaces.

The above superselection also implies that we cannot have a meaningful superposition of states that have different values of $\theta$, as they are in different Hilbert spaces~\cite{Gomez}. There is no interference or coherence between states with differing $\theta$ values. Nor would measuring $\theta$ collapse the wave function. 

Our universe thus has a specific value of $\theta$ that contributes to the electric dipole moment of the neutron through the strong CP phase 
\begin{equation}
    \bar{\theta} = \theta + Arg Det M
    \label{eq:theta-bar}
\end{equation}
where $M$ is the quark mass matrix. 

That the neutron's EDM has not yet been detected implies that $\bar{\theta}\leq 10^{-10}$ (mod $\pi)$ and its smallness compared to the CKM phase, $sin(\delta_{CKM}) \sim O(1)$, which is present in $M$ is the Strong CP puzzle.   

This is the conventional way of posing the strong CP problem, which can be resolved by axions~\cite{1977PhRvL..38.1440P, PhysRevLett.40.223,PhysRevLett.40.279} or solutions based on P and CP~\cite{Nelson:1983zb, PhysRevLett.53.329, PhysRevD.41.1286, Kuchimanchi:2010xs} 

\section{A recent approach}
However in a recent work~\cite{kaplan2025solvestrongcpproblem} the authors argue that theta eigenstates form a basis (using analogies from condensed matter systems), and thereby argue that the universe would be in an initial state that superposes states with differing $\theta$.  

Crucially, their analysis focuses only on eq.~(\ref{eq:H}) (which would allow superposing) and misses referring to eq.~(\ref{eq:O}) which splits the Hilbert space into sums of orthogonal Hilbert spaces and disallows building such superpositions.  

A simple quantum mechanical system such as a particle in a periodic potential will have theta-like eigenstates of the Hamiltonian that satisfy eq.~(\ref{eq:H}), but has observables that do not satisfy eq.~(\ref{eq:O}).   And therefore in condensed matter systems, while you can use these eigenstates as a basis to build superpositions, it is not clear that the analogy can be extended to the case of the theta vacuum of QCD.

The recent work~\cite{kaplan2025solvestrongcpproblem} also raises a concern about an argument used in the axionless solutions to the Strong CP problem where P and/or CP symmetries are imposed. In these solutions it is generally argued that since the $\theta$ term ($\sim \theta G\tilde{G}$) in the QCD Lagrangian is odd under P as well as CP (or T) operations, it would be absent (therefore in eq.~(\ref{eq:theta-bar}), the value $\theta = 0$ is chosen) if P or CP are good symmetries.  

The concern is that an analogous argument that sets $\theta$ to $0$ (or $\pi$) has not been made using the Hamiltonian formulation. We address this in the next section.

\section{Parity and the Hamiltonian}
If P is a good symmetry of the Hamiltonian then $[P,H]$= 0 and therefore all nondegenerate energy eigenstates will also be $P$ eigenstates, and since $P^2 = 1$,  these eigenstates are odd or even under parity. 

Moreover if an energy eigenstate is not parity odd or even, then it must correspond to a degenerate eigenvalue,  and the $P$ operator then relates the two degenerate states.

Under P, $\left|n\right> \ \leftrightarrow \ \left|-n\right>$ as  the winding number changes sign. We can now see using eq.~(\ref{eq:thetavac}), that the theta vacua corresponding to  $\theta = 0$ or $\pi$ are even under P. 

While in the Hilbert space with any other value of $\theta \neq 0$ or $\pi$,  the vacuum state is neither odd nor even under Parity. We can also see from ~(\ref{eq:thetavac}) that there is no second degenerate vacuum state in that same Hilbert space (with the same value of $\theta$). Therefore $P$ cannot be a good symmetry in such a Hilbert space.  

Basically for any random state in the above Hilbert space, the parity partner of the state does not exist in that same space. $P$ would take the state to a different Hilbert space with the sign of $\theta$ reversed.

Thus if P is a good symmetry, it must be the Hilbert space where $\theta = 0$ or $\pi$.

Substituting these values for $\theta$ into eq.~(\ref{eq:theta-bar}), we can see that the Strong CP phase $\bar{\theta}$ can be generated (mod $\pi$) only from $Arg Det M$ when P and CP break spontaneously through VEVs that contribute to the quark mass matrix $M$. 

The Strong CP problem can then be resolved, provided that the symmetries ensure that while the CKM phase is generated, the VEVs do not make a CP violating contribution to $\bar{\theta}$, thus explaining its observed smallness. 

Before concluding, we recall how the minimal way of restoring parity has an important experimental consequence due to the observed smallness of the strong CP phase, that can help test the model.

\section{An Experimental Consequence}

Since $P$ and $T$ (or equivalently $CP$) are discrete spacetime symmetries of the full Lorentz group, there is a strong motivation to extend the Standard Model to include them as good symmetries which are broken spontaneously.     

The simplest and most elegant way to restore $P$ is by adding three right-handed neutrinos $\nu_{iR}$ which are parity partners of the three left-handed neutrinos $\nu_{iL}$, and the $SU(2)_R$ gauge group acting on right-handed fermions with the same coupling strength as $SU(2)_L$.  

The minimal left-right symmetric model~\cite{PhysRevD.10.275,*PhysRevD.11.566,*Senjanovic:1975rk,PhysRevD.44.837, Duka:1999uc} based on $SU(3)_C \times SU(2)_L \times SU(2)_R \times U(1)_{B-L} \times P$ has a $SU(2)_L \times SU(2)_R$ bi-doublet Higgs $\phi$ that like the SM Higgs has $3\times 3$  Yukawa coupling matrices ($h^\ell, \tilde{h}^\ell$ for leptons; $h^q, \tilde{h}^q$ for quarks) that give Dirac type mass terms to the up and down sectors of the 3 generations of fermions. These matrices are Hermitian due to P.

Additionally the $SU(2)_R$ triplet Higgs ($\Delta_R)$ picks a much larger VEV $v_R$ than its $SU(2)_L$ counterpart and breaks $SU(2)_R \times U(1)_{B-L}$ to $U(1)_Y$ of the Standard Model. Its  $3 \times 3$ Yukawa matrix $f$  provides large Majorana-type masses to the right handed neutrinos and participates in the seesaw mechanism.    

The interesting thing is that since the Yukawa matrices $h^q$ and $\tilde{h}^q$ that generate the quark mass matrix $M$ are Hermitian, their determinant has no imaginary part and they do not contribute to $\bar{\theta}$ in eq.~(\ref{eq:theta-bar}).

However, the strong CP phase is generated by the imaginary part of the Higgs bi-doublet's VEV, which is determined by the imaginary part ($\alpha_{2I}$) of a dimensionless quartic coupling $\alpha_2$ of the Higgs potential, so that $\bar{\theta} \sim (m_t/m_b) \alpha_{2I}$ where  $m_t$ and $m_b$ are the top and bottom quark masses~\cite{Kuchimanchi_2015}.  

The quartic coupling $\alpha_{2I}$ is generated radiatively in RGE running of the minimal left-right symmetric model from a loop of leptons (that contributes the $Tr (f^\dagger f h^\ell \tilde{h}^\ell)$) so that~\cite{Kuchimanchi_2015}
\begin{equation}
\bar{\theta} \sim \left(1/16 \pi^2\right)(m_t/m_b)\left|Im \ Tr\left(f^\dagger f h^\ell \tilde{h}^\ell\right)\right| \ ln(\Lambda/v_R)
\label{eq:theta-oneloop-rge}
\end{equation}
where the dependence on the higher mass scales such as the P breaking scale $v_R$ and Planck scale $\Lambda$ is logarithmic.

Thus there is a fundamental connection between the Yukawa couplings of the leptonic sector, where the neutrino masses reflect the effects of spontaneous P breaking, and the strong CP phase. 

Since experimentally $|\bar{\theta}| \leq 10^{-10}$,  if CP violation also exists in the leptonic sector, it will generate too large a strong CP phase in one (and two) loop RGE running from eq.~(\ref{eq:theta-oneloop-rge}),  and therefore, assuming no fine-tuned cancellations, in most of the parameter space of the Yukawa matrices of the minimal left right symmetric model, the leptonic CP phases such as the Dirac CP phase $\delta_{CP}$ of the PMNS matrix being probed by neutrino experiments must be consistent with being $0$ or $\pi$ to within a degree~\cite{Kuchimanchi_2015}.   

Thus through neutrino experiments that will probe our expectation of $sin(\delta_{CP}) = 0$ with a high sensitivity, the Strong CP puzzle provides a way to test the left-right symmetric model in the near future, regardless of how high the $SU(2)_R$ breaking scale $v_R$ of new physics is.

We can complete the above minimal left right symmetric model in the UV by also restoring CP and adding a heavy vectorlike quark family that helps generate the CKM phase (on spontaneous breaking of CP) without generating the strong CP phase. In the minimal version of this solution~\cite{Kuchimanchi:2010xs} to the strong CP problem that restores both P and CP, leptonic CP violation is not generated and $sin(\delta_{CP}) =0$ at the tree-level with radiative corrections being negligibly small.  And the non-minimal versions are also discussed in Reference~\cite{PhysRevD.108.095023}.

The latest fit to global data from neutrino experiments by Nu-fit 6.0 (2024)~\cite{esteban2024nufit60updatedglobalanalysis} has $\delta_{CP} = \pi$ within one sigma of its error bars for normal ordering of neutrino masses. We eagerly look forward to more data from T2K and NovA, and to Hyper-K and DUNE experiments.

We also await the discovery of axions that tests the Peccei-Quinn solution, and further precision results from nEDM experiments.

\section{Conclusion}
Discrete spacetime symmetries P and T (or CP) can be imposed on the laws of physics and broken spontaneously. In this case the smallness of the strong CP phase can provide constraints on the CP violating phases of new physics and have important experimental consequences. Solutions to the strong CP problem based on P and CP symmetries can be invoked in the Hilbert space of the theta vacuum that our universe is in.

 
\bibliography{main}


\end{document}